\documentclass[12pt]{article}
\usepackage[margin=0.9in]{geometry}
\usepackage{graphicx}
\usepackage{float}
\usepackage{amsmath,amsthm,amssymb,dsfont}
\usepackage{setspace}
\usepackage{url}
\usepackage{hyperref}
\usepackage{natbib}
\usepackage{textcomp}
\setcitestyle{authoryear,open={(},close={)}}
\usepackage{xcolor} 

\usepackage{booktabs}
\usepackage{multirow}
\usepackage{multicol}

\usepackage[left, modulo]{lineno}

\newtheorem{theorem}{Theorem}
\newtheorem{prop}{Proposition}
\newtheorem{corollary}{Corollary}

\newtheorem{assumption}{Assumption}

\def\bco{\iffalse}
\hypersetup{ colorlinks=true, 
linkcolor=blue, 
citecolor=blue, 
filecolor=magenta, 
urlcolor=blue 
}

\def\T{\mathcal{T}}
\def\diff{{\rm d}}
\def\argmin#1{\mathrel{\mathop{\arg\min}\limits_{#1}}}

\def\F{Fr\'{e}chet }
\newcommand{\K}[1]{\mathrm{K}\left(#1\right)}

\def\reall{{\mathbb{R}}}
\newcommand{\E}{\mathbb{E}}

\newcommand{\iid}{\stackrel{\text{i.i.d.}}{\sim}}
\newcommand{\ind}{\stackrel{\text{ind.}}{\sim}}

\newcommand{\bla}{\color{black}}

\definecolor{purple}{RGB}{128,0,128} 

\vspace{-2cm}
\title{ \LARGE Exceedance and force of centrality for functional data}

\author{
    Poorbita Kundu$^{1}$, Hang Zhou$^{2}$, and Hans-Georg M\"uller$^{3}$ \\
    \normalsize $^{1}$Fred Hutchinson Cancer Center, Seattle, WA 98109, USA \\
    \normalsize $^{2}$Department of Statistics and Operations Research \& School of Data and Information Sciences,\\ \normalsize The University of North Carolina at Chapel Hill, NC 27599, USA \\
 \normalsize $^{3}$Department of Statistics and Data Science, University of California, Davis, CA 95616, USA \\
   }

\date{}

\begin{document}
\maketitle

\begin{abstract}
Exceedance refers to  instances where a dynamic process surpasses given thresholds, e.g.,  the occurrence of a heat wave.  
We propose a novel exceedance framework for functional data, where each observed random trajectory is transformed  into an exceedance function, which quantifies exceedance durations as a function of  threshold levels. 
An inherent relationship between exceedance functions and probability distributions makes it possible to draw on distributional data analysis techniques such as Fr\'echet regression to study the dependence of  exceedances on  Euclidean predictors, e.g., calendar year when the exceedances are observed.
We use local linear estimators to obtain exceedance functions  from discretely observed functional data with noise and study the convergence  of the proposed estimators. New concepts of interest include the force of centrality that quantifies the propensity of a system to 
revert to lower levels when a given threshold has been exceeded, conditional exceedance functions given Euclidean covariates, and threshold exceedance functions, which characterize the size of exceedance sets in dependence on covariates for any fixed threshold. We establish consistent estimation with rates of convergence for these targets. 
The practical merits of the proposed methodology are  illustrated through simulations and applications for annual  temperature curves
and medfly activity profiles.

 \vspace{1em}

\end{abstract}

\section{Introduction}
\label{sec:intro}
  In an exceedance event a dynamic process crosses a predetermined critical level or threshold.  The concept of exceedances  has been widely applied across disciplines such as environmental sciences, engineering, finance, and risk assessment. Early studies include  \citet{drufuca:77}, who modeled the total duration of exceedances for SO$_2$ concentration levels in Milan, and  \citet{graham:82}, who  applied the notion of exceedance duration to investigate wave heights. Another related perspective appeared in the early analysis literature, where \citet{hardy1930}, in the context of the maximal theorem for integrals, introduced the measure function $m(y)=\mathrm{meas}\{x : f(x)\ge y\}$, which characterizes a positive, bounded, and measurable function $f$ through the sizes of its level sets. In spatial statistics, \citet{ripley1976} developed the concept of $K$-function to describe the distribution of interpoint distances for stationary point processes
    through the behavior of neighborhood sets across different radii, with a corresponding variance-stabilizing $L$-function  \citep{besag1977,ripley1977}. Exceedance sets also play an important role in Gaussian random fields that provide a basic tool for  spatial statistics \citep{adle:76,huse:22}.\bla

In all areas of science, functional data have become increasingly prevalent.  They are usually regarded as realizations of a stochastic process  that is commonly assumed to have smooth and square integrable trajectories,  or alternatively as realizations of random elements in a Hilbert space  \citep{hsing2015theoretical}. For a comprehensive review, we refer readers to the  monographs by \cite{ramsay2005functional}, \cite{ferraty2006nonparametric}, \cite{horvath2012inference} and the review article \cite{wang2016functional}. For any function, an exceedance set consists of the elements of the domain where the function values  surpass a prespecified benchmark. When considering temperature curves, exceedance sets are of interest to quantify the prevalence of high temperatures. Similarly, for air quality data, exceedance sets can be used to assess the prevalence of high pollution episodes. \bla In engineering, exceedance sets help quantify the stress imposed on materials. Rather than the specific structure of exceedance sets,  in  many real-world scenarios the overall duration of an exceedance, i.e. the size of the exceedance sets,  is of primary interest.  \citet{chang:22} advanced the study of the size of exceedance sets by introducing occupation time curves to analyze wearable device data from the National Health and Nutrition Examination Survey (NHANES), focusing on the conversion of functional data of bounded variation to exceedance sets for a given threshold.  Our work  extends these ideas by systematically mapping functional data to  exceedance functions on a domain of thresholds and connecting these with 
distributional data analysis and associated concepts such as the force of centrality that we introduce in this paper.  

This encompasses a novel framework for functional data  that transforms random trajectories into representations that formally have the same constraints as distributions, where each trajectory is mapped to a function that quantifies the size of the exceedance sets in dependence on the  threshold level.  Instead of the original time domain, the domain of the transformed functions is an interval of thresholds. The resulting {\it exceedance functions}, where the exceedance value at a given threshold is quantified as the ratio of the Lebesgue measure of the corresponding exceedance set to the total Lebesgue measure of the domain, formally correspond to (exceedance) survival functions with associated cumulative distribution functions, and we refer to the derivatives of these distribution functions as {\it exceedance densities}, as they have the same characteristics as probability density functions. 

This perspective then leads to  the concept of  {\it force of centrality} as the ratio of the exceedance density to the exceedance survival function  at each threshold level, which formally corresponds to a hazard function in dependence on the threshold level.  In contrast to a conventional hazard function, higher values are typically preferable as they indicate reduced propensity to exceed a given threshold level. We extend these notions  to conditional versions in the presence of Euclidean covariates, such as calendar year when considering temperature exceedances. In the presence of covariates it is of interest to study {\it threshold exceedance functions}, where for a given fixed threshold the size of the exceedance sets is considered as a function of the covariate level, quantifying the propensity for an exceedance as the covariate varies. 
We develop consistent estimators for these notions when one starts with a sample of  noisy and discretely measured functional data  and utilize Fr\'echet regression \citep{petersen2019frechet} for the 
 the extension to a regression setting.  
The  paper is organized as follows. Section~\ref{sec:poptarget-1} introduces the transformation of functional data to exceedance  functions and the associated exceedance densities and force of centrality, with estimates and rates of convergence  in Section~\ref{sec:est}. Simulation results and data applications  are presented in  Sections~\ref{section:5} and~\ref{section:6}, respectively, \bla followed by a brief discussion in Section~\ref{section:7}. Proofs  and additional data illustrations are provided  in the Supplement. 

\section{Converting functional data to exceedances}

\label{sec:poptarget-1}

 Let $Y: \mathcal{T} \rightarrow \mathcal{D} \subset \mathcal{R}$ be a real-valued stochastic process defined on a compact time  domain $\mathcal{T}$ and taking values in a subset $\mathcal{D}$ of the real line $\mathcal{R}$. The \emph{exceedance set} $\Delta_Y$ at threshold value $u$ is defined as 
$$ \Delta_Y(u) = \{ t \in \T \ \text{such that} \ Y(t) \geq u\}, \ u \in \mathcal{D}.$$
For smaller  \(u\), \(\Delta_Y(u)\) is a larger subset  of the time domain,  since it is easier for \(Y(t)\) to exceed a lower threshold.
When $u$ exceeds the upper bound of the domain $\mathcal{D}$, the corresponding exceedance set is  the empty set and when it is below the lower bound it is the entire set $\T$. \bla 

The exceedance sets \(\Delta_Y(u)\) provide a localized view of when and where the process exceeds a threshold. However, in many relevant applications, only aggregated information in the form of the total size  of \(\Delta_Y(u)\) is of interest rather than the actual sets \(\Delta_Y(u)\). We focus here on such scenarios and  introduce the \emph{exceedance function} $S_Y$ as
\begin{equation}   S_Y(u) =\frac{\lambda(\Delta_Y(u))}{\lambda(\mathcal T)}, \quad  \ u \in \mathbb{R}, \label{exc} \end{equation}
where $\lambda$ denotes the Lebesgue measure and 
$S_Y$ quantifies the  relative size or duration of the exceedance event  for each  threshold \(u\). 
 Due to the stochastic nature of \( \Delta_Y(u)\), measurability of the sets   $\Delta_Y(u)$ is required to ensure that  the quantities  \( \lambda(\Delta_Y(u)) \) are well defined. 
{ We assume that the maps \((\omega,t)\mapsto Y(\omega,t)\) and \((\omega,t)\mapsto Y_i(\omega,t)\) are jointly measurable with respect to the product of the underlying probability \(\sigma\)-algebra and the Borel \(\sigma\)-algebra on \(\mathcal T\) and additionally that their sample paths are almost surely twice continuously differentiable with bounded second derivatives. Then for each fixed \(u\), the set}
$
\Delta_Y(u)(\omega) = \{ t \in \mathcal{T} : Y(\omega, t) \ge u \}
$
is a Lebesgue measurable random subset of \( \mathcal{T} \) with well defined  Lebesgue measure \( \lambda(\Delta_Y(u)) \). 
Consequently, the exceedance function
$
S_Y(u) = {\lambda(\Delta_Y(u))}/{\lambda(\mathcal{T})}
$
is a measurable stochastic process indexed by \( u \in \mathcal{D} \) and the function  \( u \mapsto S_Y(u) \) is measurable with respect to the Borel \(\sigma\)-algebra on \( \mathcal{D} \).
{ 

The function \(F_Y=1-S_Y\) is also jointly measurable. Under the  absolute continuity assumption below, \(Q_Y=F_Y^{-1}\) is measurable, and a jointly measurable version of the density \(f_Y\) is  \(f_Y(u)=\lim_{k \rightarrow \infty}k\{F_Y(u+1/k)-F_Y(u)\}\) as \(k\to\infty\); 
then \(h_Y=f_Y/S_Y\) is measurable on \(\{S_Y>0\}\).}
We note that when  normalizing the time domain of the functional data to $[0,1]$, the denominator in \eqref{exc} is not needed. The size of the exceedance sets is always constrained between 0 and 1, and \(S_Y(u)\) is a decreasing function of \(u\); higher thresholds are associated with smaller exceedance sets. \bla 

Formally, $S_Y$  has the characteristics of a survival function  and $F_Y(u) = 1 - S_Y(u)$ the characteristics of a distribution function,  which we refer to as the \emph{exceedance distribution function} of $Y$. 
{ For \(0<\lambda(\mathcal T)<\infty\), define the normalized occupation measure of a realized trajectory by \(\mu_Y(A)=\lambda\{t\in\mathcal T:Y(t)\in A\}/\lambda(\mathcal T)\), for Borel sets \(A\subset\mathbb R\). We assume  that \(\mu_Y\) is absolutely continuous with respect to Lebesgue measure  with probability one. On this event, every level set \(\{t:Y(t)=u\}\) has measure zero, so \(F_Y(u)=1-S_Y(u)=\mu_Y(( -\infty,u])=\int_{-\infty}^{u}f_Y(v)\,\diff v\), with \(f_Y=F_Y'\) almost everywhere. A sufficient condition is that almost surely \(Y\in C^1(\mathcal T)\) and \(\lambda\{t\in\mathcal T:Y'(t)=0\}=0\). }
 Analogously, we define  {\it exceedance  quantile functions}  \begin{equation}  \label{Q} {Q_Y(p)=F_Y^{-1}(p):=\inf\{u:F_Y(u)\geq p\},\quad p\in(0,1).} \end{equation}\bla
Let $\mathcal{W}=\left \{\mu\in\mathcal{P}(\mathbb{R}):\,\, \int_{\mathbb{R}}|x|^2\diff \mu(x)<\infty \right \}$    be the set of finite second moment probability measures on $\mathbb{R},$
where $\mathcal{P}(\mathbb{R})$ is the set of all probability measures on $\mathbb{R}$. {Write $\mathcal Q=\{Q_\nu:\nu\in\mathcal W\}\subset L^2([0,1])$ for the corresponding space of quantile functions.} 
For probability measures $\nu_1, \nu_2 \in \mathcal{W}$ with quantile functions $Q_{\nu_1}$ and $Q_{\nu_2}$, the 2-Wasserstein distance is given by
\begin{equation}\label{eq:dw2}
{d_W^2(\nu_1,\nu_2)=d_{L^2}^2(Q_{\nu_1},Q_{\nu_2})=\int_0^1\{Q_{\nu_1}(p)-Q_{\nu_2}(p)\}^2\,\diff p.}
\end{equation}
 Since there is  a  formal correspondence between exceedance functions and probability measures, we may quantify the distances between exceedance functions  with the 2-Wasserstein metric, which  has proved  useful in many applications 
 \citep{bolstad2003comparison, zhang2011functional}. 
 
{In the following, we will apply the distance $d_W$  to various manifestations of distributions such as densities, distribution functions, or exceedance functions to denote the distance between their underlying probability measures; densities will be featured  only when they exist. For quantile functions $Q_{\nu_1},Q_{\nu_2}\in\mathcal Q$, we use $d_{L^2}(Q_{\nu_1},Q_{\nu_2})=d_W(\nu_1,\nu_2)$.}
In survival analysis, the hazard function plays a central role in characterizing time-to-event data by describing the instantaneous risk of an event occurring at a specific time \(t\). 
of an event at \(t\), conditional on survival up to that point.  The formal analogy of exceedance functions \eqref{exc}  with survival functions  motivates the definition of the  \emph{force of centrality} on a domain of thresholds, which is a sub-domain of $\reall$ such that $F_{Y}(u)<1$,
\[
    h_Y(u) = \frac{f_Y(u)}{S_Y(u)} = \frac{f_Y(u)}{1-F_Y(u)} = -\frac{\diff}{\diff u} (\log S_Y(u)).\] 
  For a trajectory \( Y \), \( h_Y(u) \) represents the instantaneous rate at which the size of the exceedance set  $S_Y(u)$  decreases as the threshold \(u\) increases. \bla 
Larger values of  \( h_Y(u) \) indicate that the process \( Y \) has a smaller tendency to exceed thresholds beyond $u$ and thus is less likely  to exceed more extreme thresholds, while the opposite is true for smaller values of   \( h_Y(u) \).  Thus, a larger \( h_Y(u) \) means the process $Y$ is less prone to exceed more extreme thresholds, which motivates the term  ``force of centrality''. We note here that in survival analysis the survival of patients is negatively affected for larger values of a hazard function, while here a larger value is preferred, especially at thresholds $u$ where the effects of exceeding the threshold are damaging.

In summary, the exceedance function \( S_Y(u) \) serves as the foundational quantity in our framework, representing the proportion of the domain over which the process \( Y(t) \) exceeds a threshold \( u \). Its cumulative counterpart \( F_Y(u) = 1 - S_Y(u) \), the density \( f_Y(u) = \frac{\diff}{\diff u}F_Y(u) \), and the quantile function \( Q_Y(p) = F_Y^{-1}(p) \) each  characterize the exceedance distribution, where the representation through quantile functions is particularly convenient in the Wasserstein framework. Another equivalent representation of the exceedance distribution is the  force of centrality \( h_Y(u) \), which is analogous  to the hazard rate in survival analysis, but in our case captures instantaneous central tendency rather than instantaneous risk. While higher values of \( h_Y(u) \)
in a survival analysis setting are unfavorable as they indicate higher risk and lower survival,
in the context of exceedance analysis higher values of  \( h_Y(u) \) are often beneficial, as they indicate more rapid decrease of the  size of  exceedances  as  \( u \) increases; for example when studying exceedances in temperature or storm intensity.

\bla

\section{Conditional exceedance}
\label{sec:poptarget-2}

In the context of regression analysis, the data are paired observations \((X, Y)\), where \(X\) represents the covariate and \(Y\) the response variable. The primary goal is to understand the relationship between \(X\) and \(Y\). This relationship is typically characterized by the regression function \(m(x) = \mathbb{E}[Y \mid X = x]\). When \(X\) and \(Y\) involve functional data, most existing studies focus on scalar-on-function regression, where the response is a scalar and the covariates are functional data \citep{cai2006prediction, hall2007methodology, yuan2010reproducing} or on function-on-function regression \citep{yao2005aos}.

In this paper, we focus on the functional-on-vector setting, where \(Y_i\) represents functional responses and \(X_i\) one-dimensional or multivariate Euclidean predictors, and specifically on scenarios where exceedances are of primary interest. Our approach is to map the functional data to exceedance functions as previously described and to study their dependence on vector predictors. For example, in climate studies, meteorologists may examine how temperature curves over specific periods of the year vary across calendar years. 
Additionally, understanding how the sizes of exceedance sets for extreme temperatures evolve over time is of particular interest.  For instance, analyzing the relationship between calendar years and temperature exceedance functions can provide insights into changes in climate patterns. In agriculture, this analysis can reveal the number of growing days (i.e., days exceeding the minimum temperature required for plant growth) on one hand, and the number of days with damaging heat (i.e., days exceeding a temperature level that causes plant damage) on the other.

The analytical analogy of exceedance distributions and probability distributions suggests that for relating exceedance distributions with predictors we can use known tools that relate probability distributions as outcomes to vector predictors. One such tool is \F regression \citep{petersen2019frechet}, which targets the conditional expectation 
\begin{align}
\label{Q:reg:gen}
{Q_{\oplus}(\cdot,x)=\underset{q\in\mathcal Q}{\operatorname{argmin}}\ \mathbb E\left\{d_{L^2}^2(Q_Y,q)\mid X=x\right\}.}
\end{align}
Here $Q_{Y}$ denotes the exceedance quantile function associated with functional response $Y$, {$\mathcal Q$ is the quantile-function space associated with $\mathcal W$}   and $d_{L^2}$ denotes the standard $L^2$ metric, as defined in equation~\eqref{eq:dw2}. In the context of \F regression with univariate probability distributions as outcomes, the optimization problem is typically formulated in  the  space of quantile functions, {resulting in the conditional quantile exceedance function $Q_\oplus(p,x)$, where $p\in(0,1)$ is a probability level and $x$ is the predictor level. At threshold $u$, the corresponding distribution function is $F_\oplus(u,x)=Q_\oplus^{-1}(u,x)$, where the generalized inverse acts on the first argument with $x$ fixed, as defined below. When this distribution is absolutely continuous, its density is $f_\oplus(u,x)=\frac{\diff}{\diff u}F_\oplus(u,x)$ almost everywhere.} The implementation of the target \eqref{Q:reg:gen} follows an approach analogous to targeting \( \E(Y | X = x) \) in Euclidean regression by linear regression and locally linear regression, obtaining  solutions of the following optimization problems \eqref{def:GF} and \eqref{def:LF},  respectively. The target of global \F regression is 
\begin{equation}
\label{def:GF}
{Q_{\oplus,G}(\cdot,x)=\underset{q\in\mathcal Q}{\operatorname{argmin}}\ \mathbb E\left\{s_G(X,x)\,d_{L^2}^2(Q_Y,q)\right\},}
\end{equation}
where the weights in \eqref{def:GF} $ s_G(X, x):=1+(X-\mu)^T \Sigma^{-1}(x-\mu)$ are the same as those employed by linear regression when viewed as a weighted average of (Euclidean) responses. \bla For the Euclidean predictor vector $X \in \mathbb{R}^p$, we assume that the mean $\mu := \mathbb{E}(X)$ and the covariance matrix $\Sigma := \operatorname{Var}(X)$ exist, with $\Sigma$ being positive definite.

A flexible alternative  to global Fr\'{e}chet regression is local  Fr\'{e}chet regression, obtained by modifying the weight function in \eqref{def:GF} to incorporate local weights 
\citep{petersen2019frechet} 
\begin{equation}
\label{def:LF}
{Q_{\oplus,L}(\cdot,x)=\underset{q\in\mathcal Q}{\operatorname{argmin}}\ \mathbb E\left\{s_L(X,x,b)\,d_{L^2}^2(Q_Y,q)\right\}.}
\end{equation}
 Here the weights are  $s_L(z, x, b) = \frac{1}{\sigma_0^2} \mathrm{K}_b(z - x) \left[ \mu_2 - \mu_1 (z - x) \right]$, \bla where $\mathrm{K}_b(z - x) := b^{-1} \mathrm{K}((z - x)/b)$ is a kernel function $\mathrm{K}$ scaled by bandwidth $b$,  $\mu_j := \mathbb{E}[\mathrm{K}_b(X - x) (X - x)^j]$ for  $j = 0, 1, 2$, and $\sigma_0^2 = \mu_0 \mu_2 - \mu_1^2$.  An extension to low-dimensional multivariate predictors is also feasible with some straightforward modifications. Local \F regression adopts and extends the idea of local linear regression for real valued response to the general case with random object responses.  While it is more flexible, 
local \F regression is subject to the curse of dimensionality. Theoretical aspects including the existence and uniqueness of global and local Fr\'echet regression are discussed in detail in \cite{petersen2019frechet} and \cite{chen2022uniform}. \bla

The conditional distribution, density, exceedance and force functions are obtained from $Q_\oplus(p,x)$ as follows. Here $p\in(0,1)$ denotes a probability level an  $u$ a threshold, while the generalized inverse below is taken with $x$ fixed, with the convention $\sup\varnothing=0$. The density identity applies almost everywhere when the distribution is absolutely continuous, and the force of centrality $h_{\oplus}(u, x)$  is defined when $S_\oplus(u,x)>0$:
\begin{align}
    \label{def:reg:all}
    & {F_\oplus(u,x)=Q_\oplus^{-1}(u,x):=\sup\{p\in(0,1):Q_\oplus(p,x)\leq u\}}\\
    & f_{\oplus}(u, x) =\frac{\diff}{\diff u}  F_{\oplus}(u, x) \\
    & S_{\oplus}(u, x) = 1 - F_{\oplus}(u, x) = 1 - Q_{\oplus}^{-1}(u, x)\\
    & h_{\oplus}(u, x) = \frac{f_{\oplus}(u, x)}{S_{\oplus}(u, x)} = \frac{f_{\oplus}(u, x)}{1 - F_{\oplus}(u, x)}.
\end{align}

In the presence of covariates, of major interest is the size of exceedance sets as a function of covariate level $x$ for a fixed threshold $u$, as exceedances may become more or less likely with varying covariate levels.  This motivates the  definition of  the threshold exceedance function
\begin{align}
    \label{def:ref:size}
    \eta_{u}(x) =  \lambda(\mathcal{T}) \bla \ S_{\oplus}(u, x) =  \lambda(\mathcal{T}) \bla \ (1 - Q_{\oplus}^{-1}(u, x)),
\end{align}
where  $\lambda(\mathcal{T})$ \bla represents the total duration of the time domain and $S_{\oplus}(u, x)$ is the conditional exceedance function. Thus the function $\eta_{u}(x)$ quantifies the extent to which a prespecified threshold $u$ is exceeded over a given time domain in dependence on the covariate level $x$. A typical example would be the selection of a critical temperature as threshold $u$ and considering calendar year as covariate $x$. 

 While the proposed  regression approach builds upon established statistical tools such as local and global \F regression, our principal innovation lies in introducing a novel conceptual framework for functional data based on exceedance representations. This represents a distinct departure from the conventional perspective for functional data, where analysis is  carried out in the time domain with a focus on the time-dependent  features of the trajectories. By transforming functional trajectories into exceedance functions, which are distributional objects indexed by threshold levels, our framework shifts the analytical perspective from the time domain to a threshold-based domain. As functional trajectories are transformed into exceedance functions, the resulting distributional objects  characterize the proportion of time a trajectory exceeds a given threshold level. This perspective is facilitated by recent advances in distributional data analysis \citep{mata:22} and is particularly valuable for applications where the precise timing of extreme events is less relevant than their overall duration or persistence.  For example, in  studies of temperature or air pollution (e.g., daily summer temperature or PM2.5 concentration level), assessing how long a recording  remains above a critical threshold may be more relevant  than identifying the exact timings  when such exceedances occur. The exceedance representation directly captures this idea, offering an interpretable regression framework to study extreme behavior of functional trajectories. \bla

\section{Estimation and convergence}
\label{sec:est}
So far, exceedance functions, exceedance distribution functions, force of centrality  and conditional exceedance have been introduced as population-level concepts. In this section, we focus on estimating these population quantities based on observations from random samples \(\{Y_i\}_{i=1}^{n}\).  { Throughout this section, we assume without loss of generality that the time interval has been rescaled to \(\mathcal T=[0,1]\) so that \(\lambda(\mathcal T)=1\).}
Section \ref{sec:est-exce} presents the model setup, estimation procedure, and theoretical properties for the exceedance functions without Euclidean predictors.
Section \ref{sec:est-cond} then develops the estimation method and theoretical results for the conditional exceedance framework. Auxiliary results and proofs are in the Supplement.

\subsection{Exceedance functions}\label{sec:est-exce}
Although functional data theoretically consist of trajectories observed over a continuum, in practice data are usually collected or observed discretely at \(N_i\) time points with additional measurement errors.  In this common situation, the actual observations for each \(Y_i\) are given by
\[
\{(t_{ij}, Z_{ij}) \mid Z_{ij} = Y_i(t_{ij}) + \epsilon_{ij},\, j = 1, \ldots, N_i \},
\]
where the measurement errors \(\epsilon_{ij}\) are i.i.d. copies  of \(\epsilon\), with \(\mathbb{E}(\epsilon) = 0\) and \(\text{Var}(\epsilon) = \nu_0^2 < \infty\). { We further assume that the observation times \(t_{ij}\) are mutually independent across \(i\) and \(j\), and that the three collections \(\{t_{ij}\}_{i,j}\), \(\{\epsilon_{ij}\}_{i,j}\), and \(\{Y_i\}_i\) are mutually independent, which is a standard assumption in functional data analysis.} \bla  
For simplicity, we assume \(N_i = N\) for all \(i = 1, \ldots, n\) without loss of generality. 
The results can be extended to the more general scenario where the \(N_i\) vary,  with minor modifications.

Information from all subjects can be used to estimate population quantities, such as mean and covariance functions. Corresponding estimates have been extensively studied and can be constructed by  using kernel or local polynomial smoothing methods \citep{yao2005jasa, hall2006propertieshans, zhang2016sparse} or alternatively various spline-based approaches \citep{rice2001nonparametric, yao2006penalized,  cai2011optimal}.
Since exceedance functions are derived from individual trajectories, we reconstruct individual trajectories directly from the data available for each trajectory by employing local linear smoothing  \citep{mull:87:4,fan1996local}.
For $t \in \T$, the estimate for the $i$th trajectory $Y_i(t)$ is 
\begin{equation}\label{e:ll}
	\hat{Y}_{i}(t) =\argmin{\beta_{i0}\in\reall} \frac{1}{N}\sum_{j=1}^{N}\frac{1}{h}\K{\frac{t_{ij}-t}{h}}\{Z_{ij}-\beta_{i0}-\beta_{i1}(t_{ij}-t) \}^2,
\end{equation}
where $h$ denotes the bandwidth used for smoothing and K represents the kernel function. {We use notations $p\in(0,1)$ to denote probabilities that serve as arguments for quantile functions and $u$ to denote thresholds that serve as arguments of exceedance distribution, density and force functions. The covariate $x$ is held fixed when inverting or differentiating in the first argument. When this argument is omitted, as in $f_\oplus(x)$, this denotes the entire function $f_\oplus(\cdot,x)$.} \bla We require the following assumptions:
\begin{assumption}
\label{asm:bndd}
	The random functions $Y_{i}$ are positive almost surely and have a second order derivative that is bounded uniformly in $i=1,\ldots,n$. 
\end{assumption} 

\begin{assumption}
\label{asm:t_ij}
   For each \( 1 \le i \le n \), the measurement times \( t_{ij} \) form a random sample from a fixed distribution on \( \mathcal{T} \) with a density function \(f_T\) that is bounded away from 0 and \( \infty \). The functions $Y_i$ and the $t_{ij}$ are independent for $i=1,\ldots,n$ and $j=1,\ldots,N$. 
\end{assumption}

\begin{assumption}
\label{asm:K}
	The kernel function $\mathrm{K}$ is a symmetric density function defined on $[-1,1]$, and is Lipschitz continuous with Lipschitz constant $L_{K}$.
\end{assumption}

\begin{assumption}
\label{asm:rege}
	There exists $\alpha>2.5$ such that $\E|\epsilon_{ij} |^{\alpha}<\infty$.
\end{assumption}

\begin{assumption}
\label{asm:h}
   The bandwidth $h$ satisfies   ${N h^3}/{\log (n N) }\rightarrow\infty$, as $n \rightarrow \infty, \, N \rightarrow \infty$.
\end{assumption}
\noindent Assumptions \ref{asm:bndd} to \ref{asm:K} are mild conditions commonly used in kernel smoothing methods, while Assumption \ref{asm:rege} is necessary to ensure the uniform convergence of the local linear estimator \citep{zhang2016sparse}.
 Since our analysis focuses on exceedance behavior, i.e., the duration of periods during which the process exceeds a given threshold, it is natural to consider processes that take positive values. However, our results can be easily extended to processes that are not strictly positive by  adding a sufficiently large positive number so that all sample paths of \( Y \) become positive. \bla  The second-order differentiability condition in Assumption \ref{asm:bndd} refers to the sample paths of the random functions \( Y_i \) and is assumed to hold almost surely. Specifically, we assume that with probability one, each realization of \( Y_i \) is twice continuously differentiable with uniformly bounded second derivatives. This assumption ensures sufficient smoothness for the validity of the local approximations used in our analysis. \bla Assumption \ref{asm:h} specifies a rate requirement for the bandwidth \(h\) and $N$, which is critical for consistency. 
Since the exceedance function estimators rely on the reconstructed \(\hat{Y}_i\), it is crucial to study the  uniform convergence of \(|\hat{Y}_i(t) - Y_i(t)|\).

\begin{theorem}
\label{thm:Yuni}
Under Assumptions~\ref{asm:bndd}--\ref{asm:h},
\[
\sup_{1\le i\le n}\sup_{t\in\mathcal T}|\hat Y_i(t)-Y_i(t)|
=
O\!\left(
h^{2}+\sqrt{\frac{\log(nN)}{Nh}}
\right)
\qquad \text{a.s.}
\]
In particular, for \(h=\{\log(nN)/N\}^{1/5}\),
\[
\sup_{1\le i\le n}\sup_{t\in\mathcal T}|\hat Y_i(t)-Y_i(t)|
=
O\!\left(\left\{\frac{\log(nN)}{N}\right\}^{2/5}\right)
\qquad \text{a.s.}
\]
\end{theorem}

Theorem~\ref{thm:Yuni} establishes the uniform convergence of the reconstructed functions $\hat Y_i(t)$, where the convergence is  uniform  over both  time  $t \in \T$ and the subject index $i$; this is necessary  for subsequent analyses such as Fr\'echet regression.  We use \( L^2[0,1] \) to denote the space of all square-integrable functions on \([0,1]\), endowed with the inner product \( \langle g, h \rangle = \int_0^1 g(x) h(x) \, \diff x \) and corresponding norm \( \|g\|_{L^2}^2 = {\int_0^1 g^2(x) \, \diff x} \). \bla
 Under milder assumptions, the following corollary establishes the $L^2$ convergence rate for each individual trajectory $Y_i$ using standard arguments. 

\begin{corollary}\label{cor:L2Y}
	Under Assumptions~\ref{asm:bndd} to \ref{asm:K}, if the measurement errors $\epsilon_{ij}$ have finite second moments, and when choosing $h = N^{-1/5}$, it holds that  
$$\|\hat{Y}_{i}-Y_{i}\|_{{L}^2} = O_{P}\left( N^{-2/5} \right).$$
	\end{corollary}
Denote  the exceedance function  for the \(i\)th subject (defined in Section~\ref{sec:poptarget-1}) by  \(S_i := S_{Y_i}\) and  similarly for the exceedance distribution function \(F_i := F_{Y_i}\), exceedance density function \(f_i := f_{Y_i}\) and force of centrality \(h_i := h_{Y_i}\). The corresponding estimates for the $S_i$, $F_i$ and corresponding quantile functions $Q_i$ are  
\begin{align}
& \widehat{S_i}(u) = S_{\widehat{Y_i}}(u) := \lambda (\{ t \in \T \ \text{such that} \ \widehat{Y}_i(t) \geq u\}) ,\label{eq:hatS} \\
& {\widehat F_i(u)=1-\widehat S_i(u),\quad\widehat Q_i(p)=\widehat F_i^{-1}(p),\quad p\in(0,1).}\label{eq:hatF}
\end{align}
Based on Theorem \ref{thm:Yuni} and the definition of exceedance functions, the following theorem demonstrates the uniform convergence of $\hat{S}_i$, $\hat{F}_i$ and $\hat{Q}_i$, as defined in \eqref{eq:hatS} and \eqref{eq:hatF}. {As above, $u$ denotes a threshold and $p\in(0,1)$ denotes a probability level.}
\begin{theorem}\label{thm:S}
Under the assumptions of Theorem~\ref{thm:Yuni}, assume additionally that, almost surely, each exceedance distribution function \(F_i\) is absolutely continuous with density \(f_i\), and that there exists a constant \(C_f>0\) such that
\(
\sup_{i\ge 1}\|f_i\|_{\infty}\le C_f \quad \text{a.s.}
\)
Then, with
\(
b_n=\{{\log(nN)}/{N}\}^{2/5},
\)
\begin{align*}
\sup_{1\le i\le n}\sup_{u\in\mathbb R}|\hat S_i(u)-S_i(u)|
&=O(b_n)\qquad \text{a.s.},\\
\sup_{1\le i\le n}\sup_{u\in\mathbb R}|\hat F_i(u)-F_i(u)|
&=O(b_n)\qquad \text{a.s.},\\
\sup_{1\le i\le n}\sup_{p\in(0,1)}|\hat Q_i(p)-Q_i(p)|
&=O(b_n)\qquad \text{a.s.}
\end{align*}
\end{theorem}

\noindent  Let $\max|Y_i| := \max \{Y_i(t): t \in \mathcal{T}\} = \| Y_i \|_\infty$ denote the supremum (maximum) norm of the function $Y_i$. \bla To estimate the exceedance density ${f}_i$, i.e. the derivative of \(F_i\), we consider 
a positive sequence $\delta_n \rightarrow 0$ as $n \rightarrow \infty$, which we use to construct the estimate
 \( \hat{f}(u) \) as a difference quotient;  we note here that we are not actually dealing with empirical distributions but rather with functions that behave analogously. Then 
 \begin{align}
\hat{f}_i(u) = \frac{\hat{F}_i(u + \delta_n) - \hat{F}_i(u - \delta_n)}{2\delta_n}, 
\end{align}
{ which we use for all \(u\ge\delta_n\), since \(\hat F_i\) is defined on \(\mathbb R\).}
{ For \(u\in[0,\delta_n)\), we employ the forward-difference estimator}
{ \(\hat f_i(u)=\frac{\hat F_i(u+\delta_n)-\hat F_i(0)}{u+\delta_n}\).}
{The density convergence result below is restricted to \([m_i+\delta_n,M_i-\delta_n]\), where \(m_i=\min_{t\in\mathcal T}Y_i(t)\), and does not assert convergence at either endpoint of the actual support.} \bla
This  leads to the estimates for the force of centrality 
\begin{align}
  \widehat{h_i}(u) = \frac{\widehat{f_i}(u)}{1 - \widehat{F_i}(u)}.  \label{eq:hath}
\end{align} 
The following result establishes the convergence rate of $\hat{f}$ in dependence on  \( \delta_n \).
\begin{theorem}\label{thm:S'}
{Let \(m_i:=\min_{t\in\mathcal T}Y_i(t)\) and \(M_i:=\|Y_i\|_\infty\). Under the assumptions of Theorem~\ref{thm:S}, assume additionally that almost surely the restriction of \(F_i\) to its actual support \([m_i,M_i]\) is twice continuously differentiable, with derivatives at the endpoints interpreted as one-sided derivatives, and that there exists a deterministic constant \(C_2>0\) such that}
\(
{\sup_{i\ge 1}\sup_{u\in[m_i,M_i]} |F_i''(u)| \le C_2}
\) a.s.
Then, for any positive sequence \(\delta_n\to 0\),
\[
{\sup_{1\le i\le n}\sup_{u\in[m_i+\delta_n,M_i-\delta_n]}}
\left|\hat f_i(u)-f_i(u)\right|
=
O\!\left(
\frac{1}{\delta_n}\left(\frac{\log(nN)}{N}\right)^{2/5}
+\delta_n
\right)
\qquad \text{a.s.}
\]
\end{theorem}

{Selecting the optimal rate for \(\delta_n\), Corollary~\ref{cor:S'} below establishes uniform convergence of \(\hat f_i\) on the trimmed actual supports, and of the estimated force of centrality on the intersection of these transports, where the survival functions are uniformly bounded away from zero.}

{\begin{corollary}\label{cor:S'}
Under the assumptions of Theorem~\ref{thm:S'}, with the choice
\(
\delta_n=\left({\log(nN)}/{N}\right)^{1/5},
\)
\[
{\sup_{1\le i\le n}\sup_{u\in[m_i+\delta_n,M_i-\delta_n]}}
\left|\hat f_i(u)-f_i(u)\right|
=
O\!\left(\left(\frac{\log(nN)}{N}\right)^{1/5}\right)
\qquad \text{a.s.}
\]
{For any fixed \(\epsilon>0\), with}
\(
\mathcal X_{\epsilon,n}
:=
\left\{
{u\in\bigcap_{i=1}^n[m_i+\delta_n,M_i-\delta_n]:}
\min_{1\le i\le n} S_i(u)\ge \epsilon
\right\},
\)
\[
\sup_{1\le i\le n}\sup_{u\in\mathcal X_{\epsilon,n}}
|\hat h_i(u)-h_i(u)|
=
O\!\left(\left(\frac{\log(nN)}{N}\right)^{1/5}\right)
\qquad \text{a.s.}
\]
\end{corollary}}

\noindent {The sequence of sets \(\mathcal X_{\epsilon,n}\) restricts thresholds that are considered in the uniform bound  to a trimmed support that is common to all subjects  and for which also the density bound applies; and it requires that  \(S_i(u)\ge\epsilon\) for all $i$ to  keep the denominator of \(h_i=f_i/S_i\) away from zero, ensuring the $h_i$ are well defined.  A supremum over an empty evaluation set is taken to be zero.}

\subsection{Conditional exceedance}\label{sec:est-cond}

We now turn to the important conditional exceedance framework, where one has  a Euclidean \bla predictor $X$ and aims to study how exceedances depend on the values of the predictor. Observing that the responses essentially correspond to quantile functions and therefore are subject to the same restrictions as distributional data,  we adopt Fr\'echet regression 
\citep{petersen2019frechet}, a regression method for scenarios where Euclidean predictors are coupled with responses in general metric spaces, including the Wasserstein space for distributional responses. As mentioned in Section \ref{sec:poptarget-2}, one  has the choice between a local and a global version of Fr\'echet regression. 
In the following we  provide convergence results for conditional exceedance and threshold exceedance functions when using  the local version. We note that similar results can also be obtained for the global version. 

For  local Fr\'echet regression  one adopts empirical weights derived from local linear regression that are given by 
\begin{equation}
s_{in}(x, b) = \frac{1}{\widehat{\sigma}_0^2} \mathrm{K}_b(X_i - x) \left[\widehat{u}_2 - \widehat{u}_1 (X_i - x)\right],
\label{lf1} \end{equation}
where \(\widehat{u}_l = n^{-1} \sum_{i=1}^n \mathrm{K}_b(X_i - x) (X_i - x)^l\) for \(l \in \{0, 1, 2\}\), \(\widehat{\sigma}_0^2 = \widehat{u}_0 \widehat{u}_2 - \widehat{u}_1^2\), and \(\mathrm{K}_b(\cdot) = b^{-1} \mathrm{K}(\cdot / b)\)  is a kernel function $\mathrm{K}$ scaled by bandwidth $b$, typically chosen by cross-validation\bla. {The local estimator \(\widehat f_{\oplus,L}(x)\) of the conditional exceedance density \(f_\oplus(x)\) in \eqref{def:reg:all} is obtained as the density corresponding to the following conditional quantile-function estimator:}
\begin{equation} \label{lf2}
{\widehat Q_{\oplus,L}(\cdot,x)=\underset{q\in\mathcal Q}{\operatorname{argmin}}\left\|q-\frac{1}{n}\sum_{i=1}^{n}s_{in}(x,b)\widehat Q_i\right\|_{L^2([0,1])}^2,}
\end{equation}
 where $\widehat{Q}_i$ are the estimated quantile functions defined in \eqref{eq:hatS} and \eqref{eq:hatF} of Section \ref{sec:est}, obtained from the individual-level trajectory estimates $\widehat{Y}_i$. Since $\widehat{Q}_i$ are well-defined quantile functions in $L^2([0,1])$, the estimator $\widehat{Q}_{\oplus, L}$ is readily obtained by solving the optimization problem in equation \eqref{lf2}, where  the solution is given by {$\widehat Q_{\oplus,L}(\cdot,x)=\underset{q\in\mathcal Q}{\operatorname{argmin}}\ d_{L^2}^2(\hat q_x,q)$}, with $\hat{q}_x := n^{-1}\sum_{i=1}^{n}s_{in}(x,b)\widehat{Q}_i$. Note that $\hat{q}_x \in L^2[0, 1]$ and $d_{L^2}$ denotes the standard $L^2$ metric. 

For the  numerical implementation, we follow  \citet{petersen2019frechet} and use the R package \texttt{frechet} \citep{frechet}, available on CRAN. {Specifically, let $p_j$, for $j=1,\ldots,M$, denote an equispaced grid of probability levels in $(0,1)$, and define $\tilde q_j=\hat q_x(p_j)$.}
{Under Assumption~\ref{asm:lf} below, the numerical projection is instead carried out over the discretized version of \(\mathcal Q\). Thus, in addition to the monotonicity constraint \(q_1\le\dots\le q_M\), one imposes the constraints defining the restricted response space, including its discretized \(L^2\)-bound. The resulting convex projection provides a discretized approximation of the conditional quantile function {\(\widehat Q_{\oplus,L}(p,x)\)}.} An analogous approach is used to compute the global \F estimator {$\widehat Q_{\oplus,G}(p,x)$}. For further computational details, we refer the reader to Section 6.1 of \citet{petersen2019frechet}.\bla

{This leads to estimates of the  exceedance distribution function $\widehat F_{\oplus,L}(u,x)=\widehat Q_{\oplus,L}^{-1}(u,x)$, using the same generalized-inverse convention with $x$ fixed, and the exceedance function $\widehat S_{\oplus,L}(u,x)=1-\widehat F_{\oplus,L}(u,x)$. When the estimated distribution is absolutely continuous, its density is $\widehat f_{\oplus,L}(u,x)=\frac{\diff}{\diff u}\widehat F_{\oplus,L}(u,x)$ almost everywhere, with  force of centrality  $\widehat h_{\oplus,L}(u,x)=\widehat f_{\oplus,L}(u,x)/\{1-\widehat F_{\oplus,L}(u,x)\}$ where the denominator is positive. The threshold exceedance estimate is $\widehat\eta_u(x)=\lambda(\mathcal T)\{1-\widehat F_{\oplus,L}(u,x)\}$, where $\lambda(\mathcal T)$ is the total duration of the time domain and $u$ is fixed while $x$ varies.}\bla

{
To derive the convergence rate of these estimators, we use the Hilbert space representation of univariate distributions by their quantile functions and impose the following assumption.
\begin{assumption}
\label{asm:lf}
The underlying probability-measure response space \(\mathcal W\) is restricted so that its quantile-function representation  in \eqref{def:LF} and \eqref{lf2},
$
\mathcal Q=\{Q_{\nu}:\nu\in\mathcal W\}\subset L^2([0,1]),
$
is closed and convex and, for some constant \(C>0\), \(\sup_{q\in\mathcal Q}\|q\|_{L^2}\le C\).  The marginal density \(f_X\) is twice continuously differentiable on the interior of its support. Writing
$
m(x)=\mathbb E\{Q_Y\mid X=x\}\in L^2([0,1]),
$
the map \(x\mapsto m(x)\) is twice continuously Fr\'echet differentiable in the \(L^2\) sense. On every compact subset \(\mathcal X\) of the interior of the support of \(X\), \(f_X\) is bounded away from zero and \(f_X\), its first two derivatives, and \(\|m''(x)\|_{L^2}\) are uniformly bounded.
\end{assumption}

The uniform \(L^2\) bound in Assumption~\ref{asm:lf} guarantees the required Bochner integrability. Closedness and convexity ensure that the Fr\'echet estimators are uniquely defined as metric projections in \(L^2([0,1])\), and these projections are nonexpansive. The second-order \(L^2\) differentiability of \(m\) yields the usual local linear bias of order \(b^2\).
Next we establish an upper bound for the estimation error of the estimator  in \eqref{lf2}. The resulting convergence rate consists of three components: the reconstruction error of {$\widehat Q_i(p)$}, the stochastic estimation error associated with the local linear estimator, and the corresponding local linear smoothing bias.


\begin{theorem}\label{thm:lf}
Let \(x\) be a fixed interior point of the support of \(X\) such that \(f_X(x)>0\). Under the assumptions of Theorem~\ref{thm:S} and Assumption~\ref{asm:lf}, if \(b\to0\) and \(nb\to\infty\), 
\[
d_{W}\!\left(\hat{f}_{\oplus,L}(x),{f}_{\oplus}(x)\right)
=
O_{P}\!\left(
\left(\frac{\log(nN)}{N}\right)^{2/5}+b^2+(nb)^{-1/2}
\right).
\]
Consequently, if \(b\asymp n^{-1/5}\), 
\[
d_{W}\!\left(\hat{f}_{\oplus,L}(x),{f}_{\oplus}(x)\right)
=
O_{P}\!\left(
\left(\frac{\log(nN)}{N}\right)^{2/5}+n^{-2/5}
\right).
\]
\end{theorem}

The following result  gives the corresponding uniform rate. It follows directly from Hilbert-valued local linear regression and the nonexpansiveness of the projection onto \(\mathcal Q\). 
\begin{prop}\label{prop:ulf}
Let \(\mathcal X\) be a compact subset of the interior of the support of \(X\). Under the assumptions of Theorem~\ref{thm:S} and Assumption~\ref{asm:lf}, if \(b\to0\) and \(nb/\log n\to\infty\), then
\[
\sup_{x\in\mathcal X} d_W\!\left(\hat f_{\oplus,L}(x),f_{\oplus}(x)\right)
=
O_P\!\left(
\left(\frac{\log(nN)}{N}\right)^{2/5}
+b^2+\sqrt{\frac{\log n}{nb}}
\right).
\]
In particular, for \(b\asymp(\log n/n)^{1/5}\), defining
$
r_{n,N}=({\log(nN)}/{N})^{2/5}
+({\log n}/{n})^{2/5},
$
one has
$
\sup_{x\in\mathcal X} d_W\!(\hat f_{\oplus,L}(x),f_{\oplus}(x))=O_P(r_{n,N}).
$
\end{prop}
\begin{corollary}\label{cor:ulf}
Under the assumptions of Proposition~\ref{prop:ulf}, choosing  \(b\asymp(\log n/n)^{1/5}\) and assuming  that the conditional exceedance densities \(f_{\oplus}(\cdot,x)\) satisfy
$
\sup_{x\in\mathcal X}\|f_{\oplus}(\cdot,x)\|_{\infty}<\infty,
$
it holds that for each fixed threshold \(u\),
$
\sup_{x\in\mathcal X}|\hat{\eta}_{u}(x)-{\eta}_{u}(x)|=O_P(r_{n,N}^{2/3}).
$
\end{corollary}
}

To obtain global regression estimates, we adopt the empirical global weights from \citet{petersen2019frechet}, defined as 
\begin{equation}\label{gf1}
    s_{in}(x) := 1 + \left(X_i - \bar{X}\right)^T \hat{\Sigma}^{-1}(x - \bar{X}),
\end{equation}
where \(\bar{X} := n^{-1} \sum_{i=1}^n X_i\) is the sample mean and \(\hat{\Sigma} := n^{-1} \sum_{i=1}^n \left(X_i - \bar{X}\right)\left(X_i - \bar{X}\right)^T\) is the sample covariance matrix. 
We base estimates for the global Fr\'echet exceedance regression given  a Euclidean predictor on  estimated quantile functions \(\widehat{Q}_i\) in \eqref{eq:hatF}. Incorporating the empirical global weights \(s_{in}(x)\) leads to 
\begin{equation}\label{gf2}
{\widehat Q_{\oplus,G}(\cdot,x)=\underset{q\in\mathcal Q}{\operatorname{argmin}}\left\|q-\frac{1}{n}\sum_{i=1}^{n}s_{in}(x)\widehat Q_i\right\|_{L^2([0,1])}^2.}
\end{equation}
{The following result provides the convergence rate of the global Fr\'echet regression estimator; see Supplement  Section~S2 
for further details. Due to equation \eqref{eq:dw2}, the rate in Theorem \ref{thm:gf} is also the Wasserstein convergence rate for the corresponding probability measures. \begin{theorem}\label{thm:gf} Under Assumption~S1 in the Supplement,  as \(n,N\to\infty\) \newline with \(\log(nN)/N\to0\), for every fixed covariate level \(x\), the estimator in \eqref{gf2} and the global target in \eqref{def:GF} satisfy
\[
\bigl\|\widehat Q_{\oplus,G}(\cdot,x)-Q_{\oplus,G}(\cdot,x)\bigr\|_{L^2([0,1])}=O_P\!\left(\left\{\frac{\log(nN)}{N}\right\}^{2/5}+n^{-1/2}\right).
\]
\end{theorem}}

\section{Simulation Study} 
\label{section:5}

We demonstrate the applicability of the proposed methodology in recovering latent individual-level exceedance functions through simulation studies in two distinct settings, adopting  
a standard functional data generation process based on the Karhunen-Lo\`eve decomposition. {We generate functional data on a random time grid, where the observation time points $\{t_{ij}\}_{j=1}^{N}$ are drawn from a uniform distribution on the interval $(0,1)$ for each subject $i = 1, 2, \dots, n$.} The underlying trajectories are modeled as $Y_{ij} = \log{(1 + e^{V_i(t_{ij})})}$, where $V_i(t_{ij}) = \mu(t_{ij}) + \sum_{k = 1}^{K} \xi_{ik} \phi_{k}(t_{ij})$. The choices of the mean function \( \mu(t) \), orthonormal eigenfunctions \( \phi_k(t) \), and eigenvalues \( \lambda_k \) are specific to each  simulation setting and are described in more detail below. The functional principal components  are generated as  $\xi_{ik} \sim \mathcal{N}(0, \lambda_{k})$ for $k=1, \dots, K$ and $i=1, \dots, n$. To mirror real-life data and demonstrate the effect of noise variance $\nu_0$, we contaminate the true trajectories $Y_{ij}$ with measurement errors $\epsilon_{ij}$, resulting in observed values $z_{ij} = Y_{ij} + \epsilon_{ij}$. The noise specification is described separately for each simulation setting below. We evaluate the proposed approach for $N = 50, 100, 200, 500$ observations per subject and noise levels $\nu_0 = 0.05, 0.5, 1$, fixing the time interval as  $\mathcal{T} = [0, 1]$. For each $(N, \nu_0)$, the simulation study was performed for $n=200$ independent units and was repeated $B=1000$ times.

\begin{figure}[!ht]
    \begin{center}
     \includegraphics[width = 0.65\linewidth, height = 6cm]{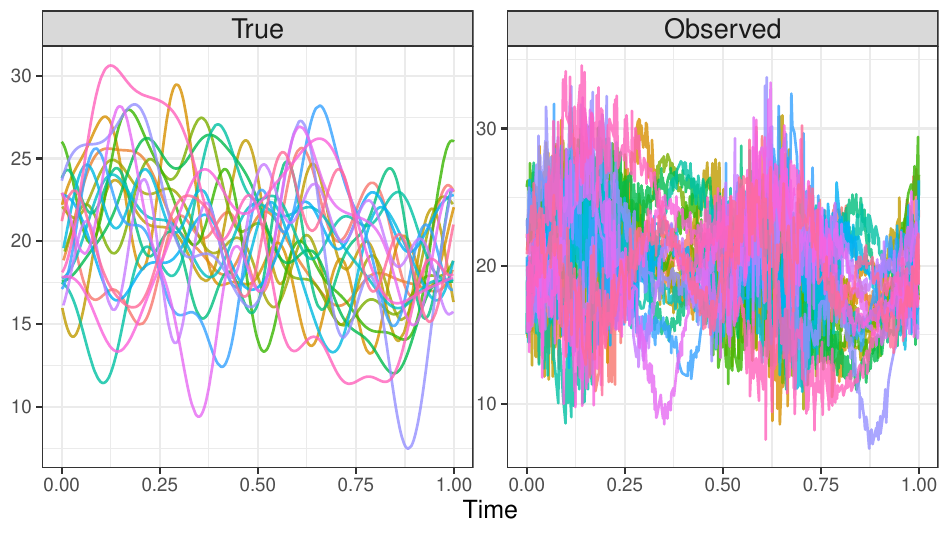}
    \caption{Simulation setting I. True underlying trajectories (left) and observed data (right) for $20$ randomly selected subjects and noise variance $1$.}
    \label{fig:Multi:Sim:Y}
    \end{center}
\end{figure}

\begin{itemize}

    \item[] \textbf{Setting I}: To model multimodal functional trajectories that exhibit high fluctuations (see Figure \ref{fig:Multi:Sim:Y}), we use the mean function $\mu(t) = m_0 + 2\sin(2\pi t) + \sin(4\pi t) + 0.5\cos(6\pi t)$ and $K = 8$ eigencomponents, incorporating higher-frequency sine and cosine components. Specifically, the eigenfunctions are $\phi_1(t) = \sqrt{2}\sin(2\pi t)$, $\phi_2(t) = \sqrt{2}\cos(2\pi t)$, $\phi_3(t) = \sqrt{2}\sin(4\pi t)$, $\phi_4(t) = \sqrt{2}\cos(6\pi t)$, $\phi_5(t) = \sqrt{2}\sin(8\pi t)$, $\phi_6(t) = \sqrt{2}\cos(10\pi t)$, $\phi_7(t) = \sqrt{2}\sin(12\pi t)$, and $\phi_8(t) = \sqrt{2}\cos(14\pi t)$, with corresponding eigenvalues $\lambda_k = c_0 k^{-a}$ for $k = 1, \dots, 8$. 
    The noise is heteroscedastic, $\epsilon_{ij} \sim N (0, \nu_0^2 (1.5 + \sin(4\pi t_{ij})))$,  with  mean function intercept $m_0 = 20$, eigenvalue decay rate $a = 1$, and overall scale $c_0 = 4$. \bla

    \item[] \textbf{Setting II}: In this relatively simpler setting,  we  consider the deterministic mean function $\mu(t) = m_0 + 2\sin(2\pi t)$ and include  $K = 3$ eigencomponents. The chosen orthonormal eigenfunctions are $\phi_1(t) = \sqrt{2}\sin(2\pi t)$, $\phi_2(t) = \sqrt{2}\cos(2\pi t)$,  and $\phi_3(t) = \sqrt{2}\sin(4\pi t)$, \bla with corresponding eigenvalues $\lambda_k = c_0 k^{-a}$ for $k = 1, 2, 3$,  with  mean function intercept $m_0 = 15$, eigenvalue decay rate $a = 1$, and overall scale $c_0 = 3$. The measurement errors are homoscedastic, with $\epsilon_{ij} \iid \mathcal{N}(0,\nu_0^2)$. \bla
\end{itemize}

We report here the simulation results for Setting I, while the corresponding illustrations for the simpler Setting II are provided in Section S3.2 of the Supplement.  For our purposes, the primary focus is on individual-level reconstruction accuracy for fixed latent trajectories. To estimate the underlying exceedance functions $S_i$, we use the estimation methodology described in \eqref{eq:hatS}. Consequently, computing the RMSE by averaging across 
$n$ subjects, so as to obtain ${\text{RMSE}_r} := {\bigl(n^{-1} \sum_{i=1}^{n} d_{W}^2(S_i, \widehat{S^r_i})\bigr)}^{1/2}$ for the $r$th replicate, would
not accurately reflect estimation error patterns at the individual level. Instead, to assess reconstruction accuracy for individual latent trajectories, we adopt the following approach: for each subject, the true exceedance function \( S_i \) is fixed across all simulation replications, while the observed data are generated randomly according to the specified simulation settings (with varying \( N \) and \( \nu_0 \)). The subject-level exceedance functions are then estimated based on the observed data in each replicate. For the $r$th replicate with estimate $\widehat{S^r_i}$, the error is quantified as $d_{W}^2(S_i, \widehat{S^r_i})$ based on the corresponding metric defined in Section~\ref{sec:poptarget-1}. Averaging over $B = 1000$ simulated datasets, the individual-level root-mean-square errors are given by ${\text{RMSE}_i} := {\bigl(B^{-1} \sum_{r=1}^{B} d_{W}^2(S_i, \widehat{S^r_i})\bigr)}^{1/2}$. Thus, this design quantifies reconstruction error. 

\bla Table \ref{tab:subject_rmse_summary}
presents the mean error ${\overline{\text{RMSE}}} := n^{-1} \sum_{i=1}^{n} {\text{RMSE}_i}$, illustrating how errors vary with different values of $N$ and noise levels $\nu_0$.  Since the absolute magnitude of estimation errors is inherently influenced by the specific choices of the data-generating mechanism and associated constants, our focus is on illustrating comparative trends in estimation accuracy as a function of \( N \) and \( \nu_0 \). Importantly, in our setting, the scale of the true subject-specific exceedance functions $S_i$ is held fixed across all replications, ensuring that RMSE$_i$ values are directly comparable across different values of $N$ and $\nu_0$.  This makes it possible 
to reliably capture relative changes in estimation accuracy driven by sample size and noise level, without being unduly affected by scale variation. 
For  applications where the scale may vary, reporting a scale-normalized error (relative to the true exceedance function) is recommended. 
The results in Table \ref{tab:subject_rmse_summary} show that the estimation errors (averaged over $n= 200$ subjects) keep decreasing as the number of observations per subject increases, also reflecting changes in noise levels. \bla

\begin{table*}[!ht]
    \caption{Individual-level RMSE averaged over $1000$ Monte Carlo runs for the estimation of exceedance functions (defined in Section \ref{sec:est-exce}) in simulation setting I for $n=200$ subjects with varying number of observations $N$ available per subject and varying noise level $\nu_0$. The reported values denote the mean RMSE values averaged over $200$ subjects.}
    \label{tab:subject_rmse_summary}
    \setlength{\tabcolsep}{2pt} 
    \begin{tabular*}{\textwidth}{@{\extracolsep{\fill}}lccc}
        \toprule
        \multirow{2}{*}{\textbf{\(N\)}} & \multicolumn{3}{c}{\textbf{Noise Level} \(\nu_0\)} \\
        \cmidrule(lr){2-4}
         & 1.0 & 0.5 & 0.05 \\
        \midrule
        50 &  0.995 & 0.957 & 0.943 \\
        100 & 0.793 & 0.764 & 0.754 \\
        200 & 0.402 & 0.358 & 0.326 \\
        500 & 0.212 & 0.126 & 0.069 \\
        \bottomrule
    \end{tabular*}
\end{table*}

To assess the finite sample performance of global Fr\'{e}chet regression estimates, we use the following data generating mechanism. We generate a one-dimensional covariate as $X \sim U(a_0, b_0)$ and $\kappa(x) = a_1 + b_1 x + \epsilon_1$, where $\epsilon_1 \sim \mathcal{N}(0, 0.5)$. We use the mean function $\mu(t, x) = \kappa(x) \mu(t) $, where $\mu(t)$ is as defined above. With the same orthonormal basis functions $\phi_k$ and eigenvalues $\lambda_k$ as before, we generate the functional principal components $\xi_{ik} \ind \mathcal{N}(0,\lambda_{k})$ for each eigencomponent (indexed by $k = 1, \dots, K$). The individual trajectories are thus generated using  a  Karhunen-Lo\`eve expansion, with white noise $\epsilon_{ij} \iid \mathcal{N}(0,\nu_0^2)$. We choose $m_0 = 2, a_0 = 1, b_0 = 5, a_1 = 1, b_1 = 5$.  From the simulated functional data, we compute the ``oracle" conditional exceedance functions (see Section~\ref{sec:poptarget-2} for definition) analytically from the true conditional distribution of the latent (noise-free) functional trajectories given $X = x$. This follows standard practice in simulation studies for functional regression, where these oracle functions (known from the data-generating model used in our simulation setting) serve as idealized targets and benchmarks for evaluating estimator performance. 
Corresponding estimated exceedance functions are obtained via global Fr\'{e}chet regression with  predictor $X$ (see Section~\ref{sec:est} for details). \bla A similar generation method is used for local Fr\'{e}chet regression, where the continuous covariate $X$ is generated from a truncated normal $(\mu_X, \sigma_X)$ with support $[l_X, u_X]$. We further use the non-linear power function $\kappa(x) = a_1 + b_1 x+ c_1 x^{3/2} + \epsilon_1$, where we choose $\mu_X = 3, \sigma_X = 0.5, l_X = 1, u_X = 5, a_1 = 10, b_1 = 5, c_1 = 2, m_0 = 4$. 

\begin{figure}[!ht]
    \begin{center}
    \includegraphics[width = 0.7\linewidth, height = 8.5cm]{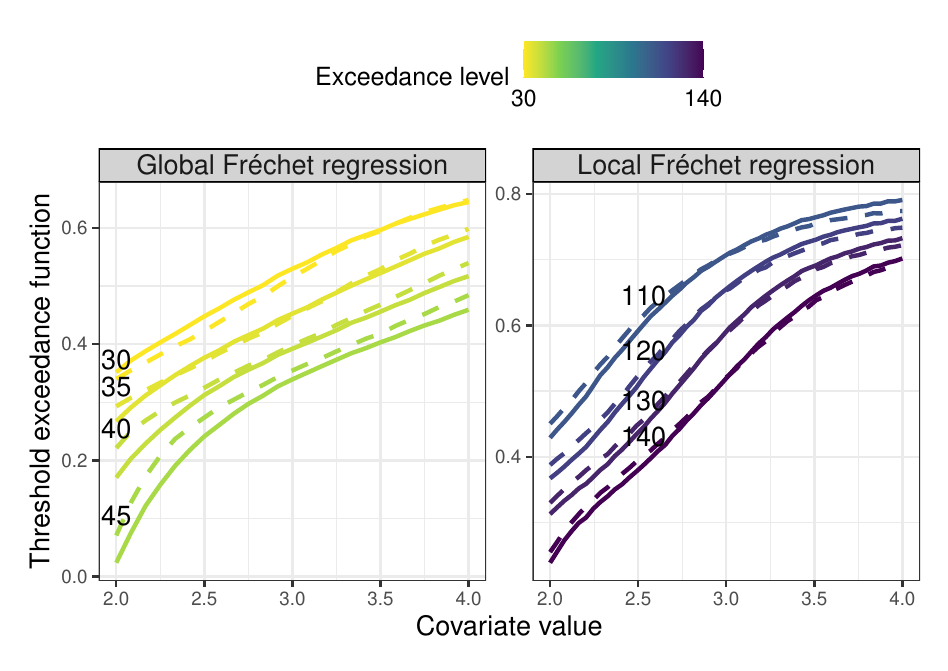}
    \caption{Conditional global \eqref{gf1},\eqref{gf2} (left) and local \eqref{lf1}, \eqref{lf2} (right) Fr\'{e}chet regression functions for one simulation run in simulation setting I. Threshold exceedance functions (response) at varying exceedance levels as a function of covariate value $x$. The dashed lines represent the ``oracle'' threshold exceedance functions and the solid lines depict their estimated counterparts.} 
    \label{fig:Multi:Sim:GloLocDenReg:size}
    \end{center}
\end{figure}

 The boundedness condition in Assumption~\ref{asm:lf} is imposed as a sufficient regularity condition for the asymptotic theory, while the numerical experiments use the standard Gaussian FPC-score construction commonly employed in functional data analysis. Analogous to the global case, the ``oracle" local Fr\'{e}chet regression functions are obtained from the true conditional distribution, with corresponding estimates obtained from simulated data based on the fitted local \F regression model. \bla Figure ~\ref{fig:Multi:Sim:GloLocDenReg:size} illustrates results for simulation Setting I  for one run over a dense grid of covariate values for $n = 50$, $N = 50$ and noise level $\nu_0 = 1$. 
We observe close alignment between the oracle and estimated threshold exceedance functions (the corresponding conditional exceedance functions are in the Supplement Figures S10 and S11). 

We also evaluated the out-of-sample error, quantified by the previously defined root-mean-square error (RMSE), for the conditional exceedance functions obtained from both global and local \F regression.  In each simulation run, the \F regression models are fitted using a training set of simulated functional trajectories and their associated covariates. To assess predictive performance, we obtain the estimated conditional exceedance functions $\widehat{S}_{\oplus,G}(x)$ (based on the fitted model) at covariate values that are excluded from the training data, thereby mimicking an out-of-sample prediction setting. Prediction accuracy is  assessed by comparing these estimated (predicted) exceedance functions with their oracle counterparts ${S}_{\oplus,G}(x)$. For the $j$th out-of-sample prediction at covariate value $x_j$, the error is quantified as $d_{W}^2\bigl({S}_{\oplus,G}(x_j), \widehat{S}_{\oplus,G}(x_j)\bigr)$, based on the corresponding metric defined in Section~\ref{sec:poptarget-1}. For each Monte Carlo simulation run, the error is quantified using the root-mean-square error (RMSE), defined for the $r$th replicate as ${\text{RMSE}_r} := {\bigl(D^{-1} \sum_{j=1}^{D} d_{W}^2\bigl({S}_{\oplus,G}(x_j), \widehat{S}_{\oplus,G}(x_j)\bigr)}^{1/2}$, where the average is taken over a grid of $D$ out-of-sample covariate values. \bla Table~\ref{tab:Multi:Frechet:RMSE:n} summarizes the mean and standard deviation of the RMSE across $500$ Monte Carlo simulations for various combinations of sample size $n$ and number of observations per subject $N$.  As expected, RMSE consistently decreases with increasing $N$, and for fixed $N$, its  variability 
generally decreases as sample size $n$ increases. \bla

\begin{table}[htbp]
  \centering
  \caption{RMSE for conditional exceedance functions obtained from global and local \F regression models (see Sections \ref{sec:poptarget-2} and \ref{section:5}) for simulation setting I with varying number of subjects $n$ and different number of observations $N$ available per subject, at noise level $\nu_0 = 1$. The values indicate the mean RMSE values, averaged over $500$ simulation runs, with respective standard deviations in the parentheses.}
  \label{tab:Multi:Frechet:RMSE:n}
\begin{tabular*}{\textwidth}{@{\extracolsep{\fill}}lccccc}
    \toprule
    \multirow{2}{*}{\textbf{Model}} & \multirow{2}{*}{\textbf{$n$}} & \multicolumn{4}{c}{\textbf{$N$}} \\
    \cmidrule(lr){3-6}
     &  & 50 & 100 & 200 & 500 \\
    \midrule
    \multirow{3}{*}{\textbf{Global}} 
      & 50  & 2.820 (0.473) & 2.086 (0.418) & 1.233 (0.330) & 0.957 (0.251) \\ 
      & 100 & 2.767 (0.403) & 2.049 (0.408) & 1.189 (0.325) & 0.893 (0.223) \\ 
      & 200 & 2.712 (0.414) & 2.003 (0.362) & 1.152 (0.307) & 0.861 (0.216) \\ 
    \midrule
    \multirow{3}{*}{\textbf{Local}}
      & 50 & 8.993 (1.078) & 6.189 (0.959) & 3.890 (1.053) & 2.150 (0.611) \\ 
      & 100 & 8.356 (0.715) & 5.480 (0.583) & 3.068 (0.963) & 1.526 (0.451) \\ 
      & 200 & 7.956 (0.459) & 5.135 (0.359) & 2.373 (0.840) & 1.169 (0.563) \\ 
    \bottomrule
\end{tabular*}

\end{table}

\section{Data Illustrations}
\label{section:6}

\subsection{Summer Temperature Data}
\label{data:summer:temp}
\begin{figure}[!ht]
    \begin{center}
    \includegraphics[width = 0.7\linewidth, height = 11cm]{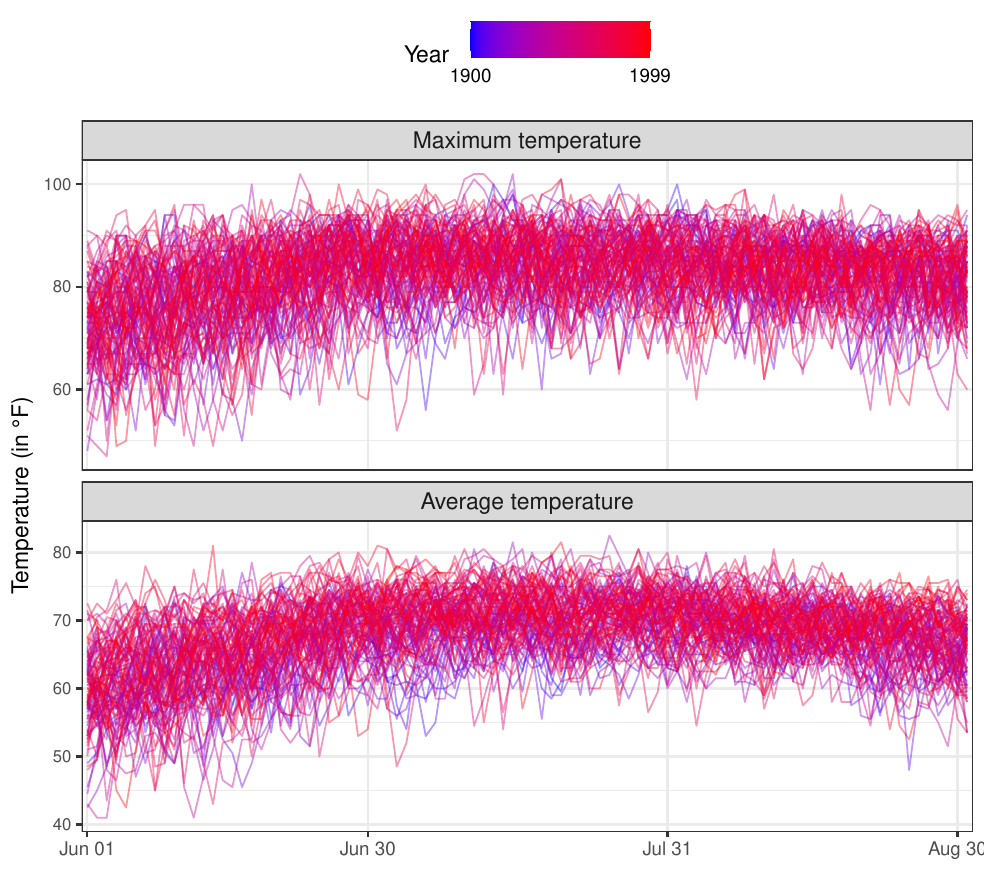}
    \caption{Daily summer temperature data (measured in degrees Fahrenheit) recorded at Fort Collins (Colorado) in the months of June-August, for the years 1900 to 1999, where years are colored blue to red. Maximum temperature (top panel) and average temperature (bottom panel). } 
    \label{fig:data:temp:FC:obs}
    \end{center}
\end{figure}

 To illustrate the real-world application of the proposed exceedance framework, we analyze temperature data recorded in Fort Collins, Colorado, for the summer months of June through August, which constitute the time domain $\mathcal{T}$,  for the years 1900 to 1999. The data are obtained from the \texttt{FCwx} dataset in the \texttt{extRemes} package \citep{extRemes:16} and provide daily maximum and minimum temperatures measured in degrees Fahrenheit. For each day, we define the average daily temperature as $(T_{\max}+T_{\min})/2$. For each year, the observed daily temperatures are viewed as discrete measurements of an underlying continuous temperature process, observed on a regular 24-hour grid, with one observation per day over the three summer months (see Figure~\ref{fig:data:temp:FC:obs}). Especially for temperature data,   extremes and the associated exceedances  are crucial for understanding changes in climate patterns and their potential impacts \citep{burke:15, ebi:21}, motivating the  study of changes in   maximum and average summer daily temperatures by converting temperature profiles to exceedances. \bla 

 In the context of the summer temperature study, the exceedance set for a particular year at a given temperature threshold comprises the summer days (June-August) during which the maximum (or average) daily temperature exceeds the specified threshold. In many practical settings, one is less interested in the specific days or times when an  exceedance occurs, but rather in 
the overall duration of exceedance events.  This motivates the use of  exceedance functions for temperature profiles,  which quantify the relative duration of exceedances  at each temperature threshold and the study of the {force of centrality} and of  conditional and threshold exceedance functions. The  {force of centrality} offers a more refined perspective by characterizing the instantaneous rate of decline of the exceedance function with  increasing temperature thresholds. Intuitively, the force of centrality at a given threshold quantifies how rapidly the occurrence of exceedance events diminishes as the threshold becomes more extreme. At a given temperature threshold, higher values of the  force of centrality indicate a reduced tendency to exceed more extreme temperatures and therefore are less concerning than lower values -- this is in contrast to the force of mortality in survival analysis where higher values indicate increased risk. Therefore, studying the evolution of the force of centrality as a function of calendar year illustrates how the tendency towards larger exceedances varies with time and reflects a crucial feature of  long-term temperature trends, particularly at temperatures  where exceedances are associated with adverse impacts ranging from agriculture to health and mortality. \bla

\begin{figure}[!ht]
    \begin{center}
    \includegraphics[width = 0.7\linewidth, height = 12cm]{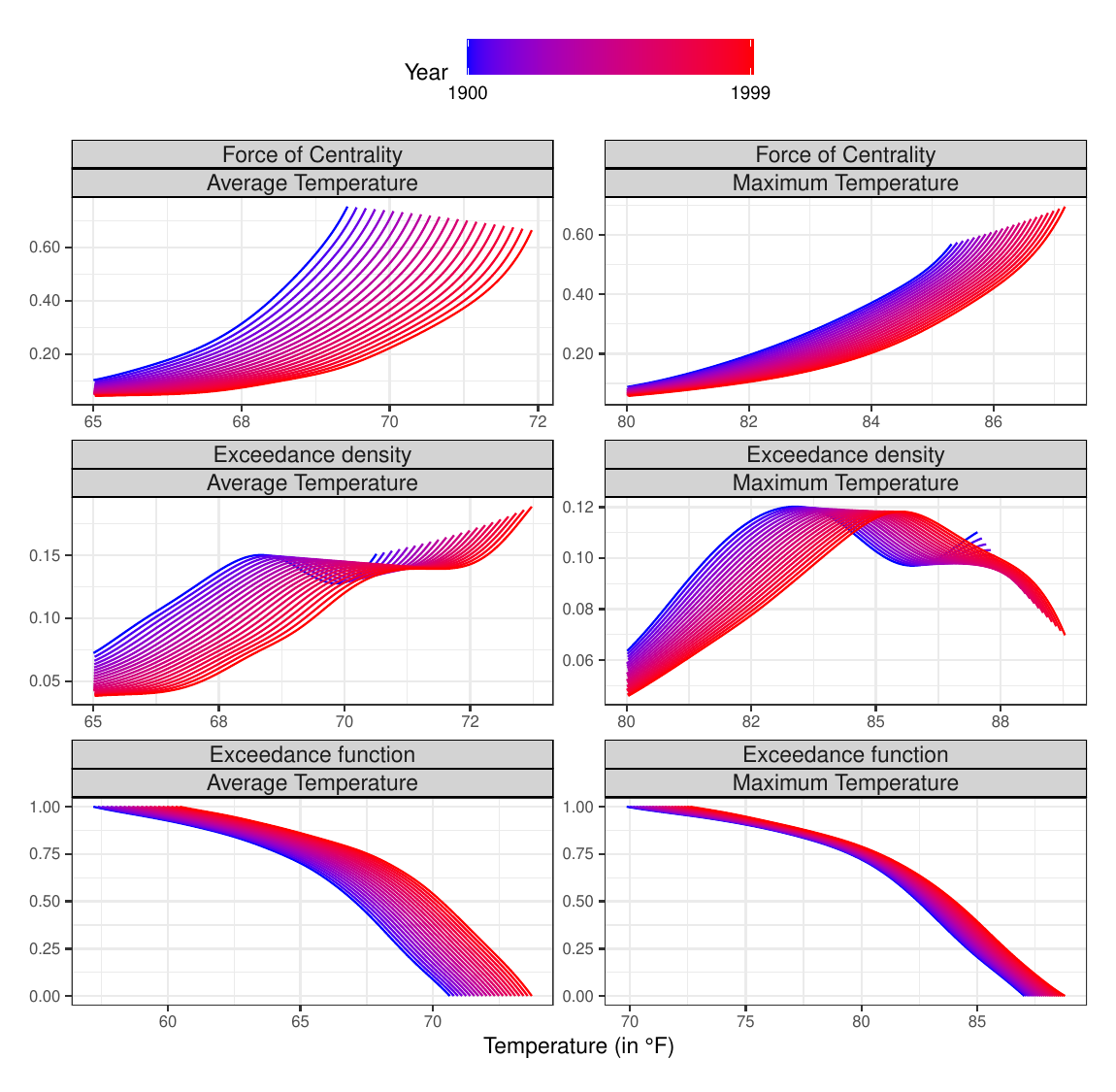}
    \caption{Daily summer temperature data for Fort Collins (Colorado) for $1900-1999$, with temperature measured in degrees Fahrenheit viewed as exceedance levels. Conditional force of centrality (top panel), exceedance densities (middle panel) and exceedance functions (bottom panel) in dependence on calendar year for average temperature (left) and maximum temperature (right), obtained with global Fr\'{e}chet regression  (see \eqref{gf1} and \eqref{gf2}).} 
    \label{fig:data:temp:3}
    \end{center}
\end{figure}

We fit a Fr\'{e}chet regression model with year as predictor and estimated summer temperature exceedance densities   and associated forces of centrality \bla as functional response. Figure~\ref{fig:data:temp:3} demonstrates a rightward shift in the support (and peaks) of the conditional exceedance densities, suggesting a trend of increasing summer temperatures \citep{dav:23}. The conditional force of centrality reflects a similar pattern: For earlier years,  the force of centrality tends to rise steeply to assume higher values over a narrower range of temperatures, indicating comparatively cooler summers in the past. The conditional exceedance functions in the bottom panels further illustrate the varying size of the exceedance sets. For each fixed temperature, this function indicates the size of the corresponding exceedance sets, normalized to a fraction between 0 and 1. With increasing calendar years, the size of the exceedance sets is found to increase over the displayed temperature range. 

\begin{figure}[!ht]
    \begin{center}
    \includegraphics[width = 0.7\linewidth, height = 7cm]{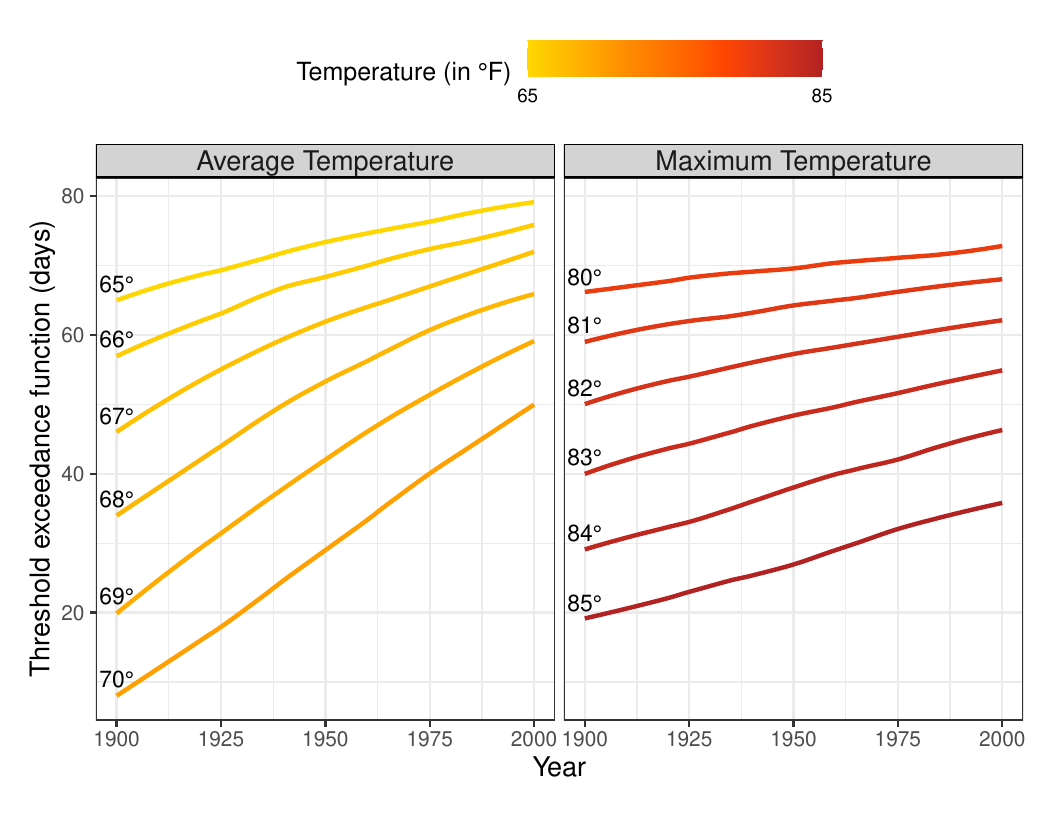}
    \caption{Conditional threshold exceedance functions at different exceedance levels (temperature measured in degrees Fahrenheit) as a function of year for average temperature (left) and maximum temperature (right), based on fitted global \F regression model for the daily summer temperature data for Fort Collins.} 
    \label{fig:data:temp:size}
    \end{center}
\end{figure}

Threshold exceedance functions provide an effective tool to demonstrate temperature trends over time  by tracking the annual number of summer days that surpass a specified temperature threshold. As illustrated in Figure \ref{fig:data:temp:size}, for an average daily temperature of 70\textdegree F, the threshold exceedance function increases from less than 10 days in 1900 to 50 days in 1999, a very substantial increase.  Similarly, fixing the maximum daily temperature at 85\textdegree F, the threshold exceedance function shows an increase from less than 20 days to more than 35 days.  Further results incorporating ENSO (El~Ni\~no--Southern Oscillation) as an additional covariate in a global-scale temperature study are provided in Section S3.3 of the Supplement. \bla 


\bco

\subsection{PM2.5 Pollution Data}
\label{data:PM}

We apply our methods for PM2.5 (fine particulate matter) levels for $242$ U.S. cities across different  states in the year $2022$ (see Figure~\ref{fig:data:PM:obs}). PM2.5, which consists of particles less than $2.5$ micrometers in diameter, is a key indicator of air pollution with significant impact for respiratory and cardiovascular health \citep{crouse2012risk, wu2020evaluating}. The PM2.5 values, measured in $\mu g/m^3$, are viewed as exceedance levels. We focus on  PM2.5 levels in relation to economic and environmental factors such as Gross Domestic Product (GDP) per capita and climatic regions.

\begin{figure}[!ht]
    \centering
\includegraphics[width=0.95\linewidth, height = 10cm]{Plots/Obs_PM_10random.pdf}
    \caption{PM2.5 pollution data (measured in $\mu g/m^3$) for $10$ randomly selected U.S. cities in the year $2022$, as reported by the United States Environmental Protection Agency.}
    \label{fig:data:PM:obs}
\end{figure}

Using state-level GDP per capita as an indicator of economic development, we first investigate the impact of economic factors on air pollution \citep{vandenbroucke2017effects, dang2020does}. We fit a global Fr\'{e}chet regression model with GDP per capita as predictor and (estimated) force  of centrality as response. To account for the role of geographical and meteorological factors in dispersion and accumulation of particulate matter, we include US climatic regions (see \url{https://www.ncei.noaa.gov/access/monitoring/reference-maps/us-climate-regions}) as an additional predictor  \citep{tai2010correlations}. 

The conditional force of centrality at varying quantiles of GDP are estimated for two climatic regions: the Ohio Valley and the South. Recall that a smaller force of centrality at a given threshold suggests a higher instantaneous chance of exceeding the threshold further. Given that a 24-hour PM2.5 concentration of $9 \ \mu g/m^3$ is classified as ``good'' by the National Ambient Air Quality Standards (NAAQS), our primary focus is on exceedance levels beyond this benchmark. As illustrated in Figure~\ref{fig:data:PM2.5:1}, the two climatic regions exhibit a broadly similar pattern. With increase in GDP, the force of centrality tends to assume lower values while extending to higher PM2.5 levels beyond $9 \ \mu g/m^3$. This suggests that regions with higher GDP per capita generally experience elevated PM2.5 levels, highlighting a potential trade-off between economic growth and environmental sustainability \citep{liu2020heterogeneous, an2021impact}. However, the conditional force of centrality attains higher values and extend to lower PM2.5 levels in the Ohio Valley climatic region. This indicates that, adjusting for GDP per capita, the tendency to exceed a given PM2.5 level above $9 \ \mu g/m^3$ is higher for cities in the South climatic region and lower in the Ohio Valley region. 

\begin{figure}[!ht]
    \begin{center}
    \includegraphics[width = 0.75\linewidth, height = 9.5cm]{Plots/fig5PM2_v2.pdf}
    \caption{PM2.5 pollution data for $242$ U.S. cities in the year $2022$, with PM2.5 values measured in $\mu g/m^3$ viewed as exceedance levels. Conditional force of centrality as a function of GDP per capita for cities in US climatic regions Ohio Valley and South, based on fitted global Fr\'{e}chet regression model.} 
    \label{fig:data:PM2.5:1}
    \end{center}
\end{figure}

Figure \ref{fig:data:PM2.5:1:size} illustrates a consistent pattern in the conditional threshold exceedance functions, showing variations in PM2.5 levels across different ranges within the three climatic regions. In the Ohio Valley, the annual number of days exceeding a PM2.5 concentration of $10.5 \ \mu g/m^3$ increases from approximately 10 days in cities with the lowest GDP to 30 days in those with the highest GDP. A similar trend is observed in the South, where the threshold exceedance function for PM2.5 level $11 \ \mu g/m^3$ increases by almost 40 days between the lowest and highest GDP cities.

\begin{figure}[!ht]
    \begin{center}
    \includegraphics[width = 0.75\linewidth, height = 10.5cm]{Plots/fig5PM2_size_v2.pdf}
    \caption{PM2.5 pollution data for $242$ U.S. cities in the year $2022$, with PM2.5 values measured in $\mu g/m^3$ viewed as exceedance levels. Conditional threshold exceedance functions at different PM2.5 values as a function of GDP per capita for cities in US climatic regions Ohio Valley, and South, based on fitted global Fr\'{e}chet regression model.} 
    \label{fig:data:PM2.5:1:size}
    \end{center}
\end{figure}

\fi

\subsection{Medfly Activity Profile Data}
\label{data:medfly}

The analysis of activity profiles and their association with diet and longevity has emerged as a significant area of research in biological and life sciences \citep{tabac:20, schr:23, chen:24, iao:24}. We illustrate our proposed method for medfly activity profile data, with lifetime activity profiles available for $96$ female Mediterranean fruit flies (medflies). Each of them was fed with one of  three agar-based gel diets, differing in their sugar and yeast hydrolysate content (50\%, 20\% and 10\%, represented by treatment C50, C20 and C10 accordingly); every diet group comprised $32$ medflies. Each fly was placed in its own glass tube with three infrared light monitors at three different locations. Daily locomotory activity count, defined as the total number of times a fly passed through the middle infrared beam, was recorded every day for each medfly until the death of all medflies (so there is no data censoring), using  Monitor-LAM25 systems. Further details on the experimental setup can be found in \citet{chen:24}.

\begin{figure}[!ht]
    \centering
\includegraphics[width=0.65\linewidth, height = 6cm]{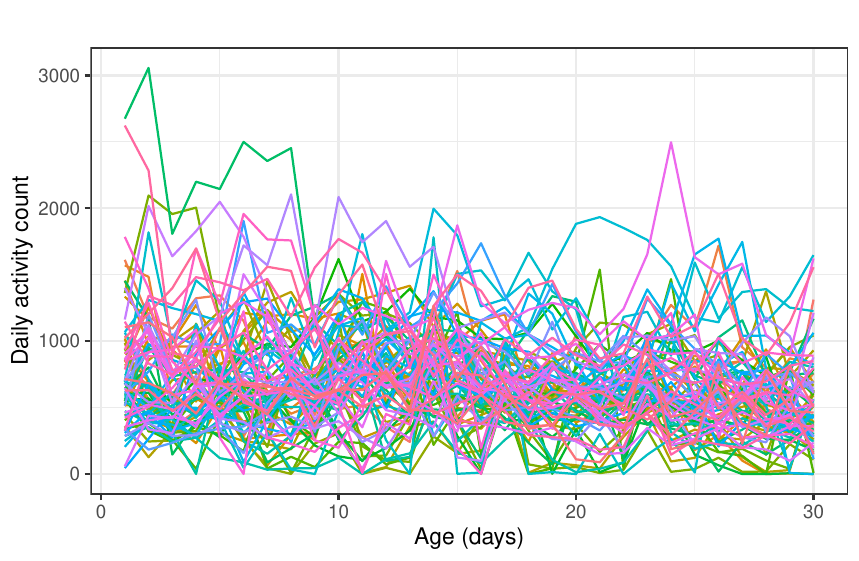}
    \caption{Individual daily activity counts in the first $30$ days after eclosion for $80$ Mediterranean fruit flies surviving through age $30$ in the medfly activity profile study.}
    \label{figMedflyObs}
\end{figure}

\begin{figure}[!ht]
    \begin{center}
    \includegraphics[width = 0.7\linewidth, height = 7cm]{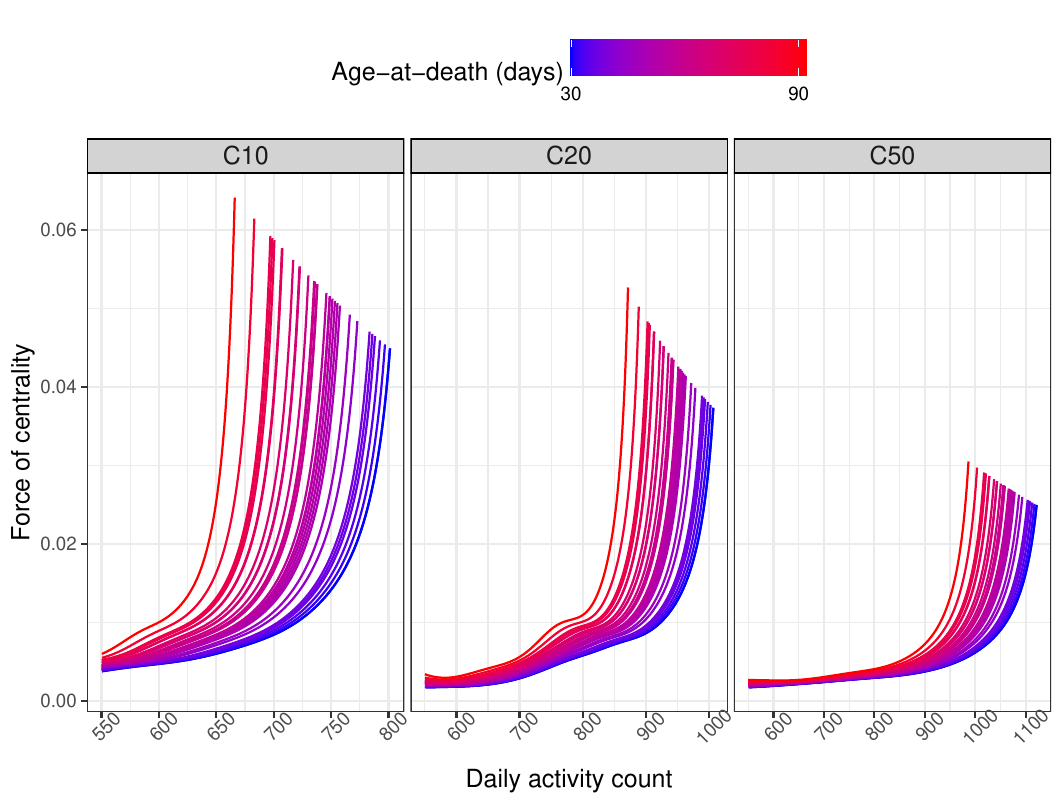}
    \caption{Medfly activity profile study with daily locomotory activity counts viewed as exceedance levels. Conditional force of centrality  as a function of age-at-death for the three diet groups with varying glucose-protein concentration: C10 (10$\%$), C20 (20$\%$), and C50 (50$\%$), based on fitted global Fr\'{e}chet regression model.
    } 
    \label{fig:data:medfly:1}
    \end{center}
\end{figure}

We focus on medflies that survive through a common age window of $\left[0, 30\right]$ days and study their daily locomotory activity in the first $30$ days of life, presented in Figure~\ref{figMedflyObs}. To study the association between activity count and lifespan, quantified as age-at-death, we fit a global Fr\'{e}chet regression model with estimated force of centrality  as  response and age-at-death as predictor. We further include diet group as an additional predictor. 
The conditional force of centrality at varying quantiles of age-at-death is depicted in Figure~\ref{fig:data:medfly:1} for each diet group. Within each diet group, longer lived medflies with higher age-at-death exhibit a force of centrality that rises steeply to higher values over a narrower range of daily activity counts. This illustrates that flies who survive longer are less likely to exceed a given activity count, thereby indicating reduced overall activity in the first 30 days of life for longer-lived flies.  The finding aligns with the broader biological concept of trade-offs between energy expenditure and longevity in insects \citep{chen:24, iao:24}. Although the overall patterns remain consistent, there are distinct differences in daily activity across diet groups. Specifically, as the sugar-yeast hydrolysate content in the diet increases, the force of centrality tends to take lower values and extends to higher activity levels. This suggests that medflies in the diet group C50 are the most active, followed by C20, and then C10. 

\begin{figure}[!ht]
    \begin{center}
    \includegraphics[width = 0.7\linewidth, height = 8cm]{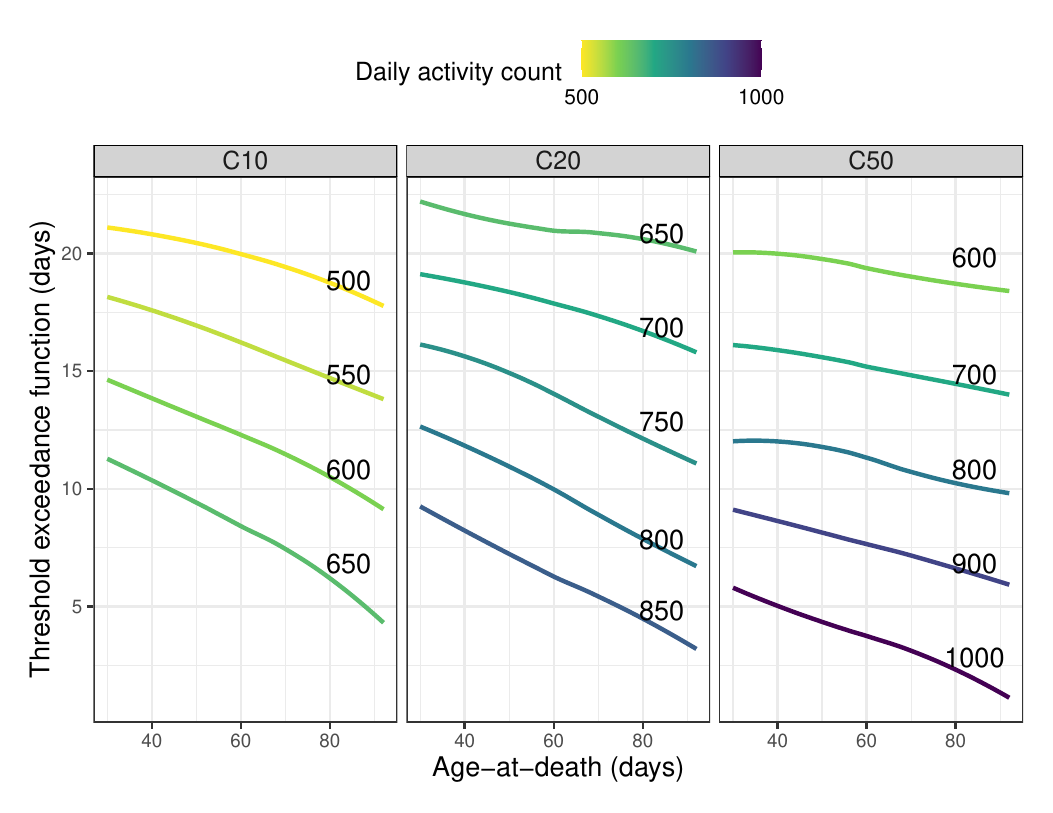}
    \caption{Medfly activity profile study with daily locomotory activity counts viewed as exceedance levels. Conditional threshold exceedance functions at varying activity counts as a function of age-at-death for the three diet groups with varying glucose-protein concentration: C10 (10$\%$), C20 (20$\%$), and C50 (50$\%$), based on fitted global Fr\'{e}chet regression model.
    } 
    \label{fig:data:medfly:1:size}
    \end{center}
\end{figure}

A similar pattern is found for the threshold exceedance functions shown in Figure~\ref{fig:data:medfly:1:size}. In diet group C10, the number of days (within the first 30 days of life) on which the activity count exceeds $650$ declines from $10$ days for flies with a lifespan of $40$ to less than $5$ days for those surviving until $90$. In C20-fed medflies, for an activity count threshold of $850$, this decrease is more than $5$ days. Among flies fed with C50, at an activity count threshold of $1000$, the number of exceedance days decreases from $5$ days at an age-at-death of $40$ to nearly zero at an age-at-death of $90$.

\subsection{NHANES accelerometer data}
\label{data:nhanes}

We further demonstrate the exceedance approach for  minute-level accelerometer data from the National Health and Nutrition Examination Survey (NHANES) for approximately $12,000$ participants. Viewing physical activity counts as exceedance levels, we examine their relationship with various demographic and socioeconomic covariates, such as age, sex, Body Mass Index (BMI), and Poverty Income Ratio (PIR), with details in Section S3.1 of the Supplement. 

\section{Discussion}
\label{section:7}

The proposed exceedance  approach  that transforms random trajectories into exceedance functions with their formal analogy with survival functions provides a new tool to  study extreme events that occur for dynamic processes within the broader context of functional data. This nonparametric methodology circumvents the challenges of model misspecification associated with traditional distributional exceedance modeling and  is underpinned by  theory, including uniform convergence rates for individual-level estimates.
We introduce the force of centrality that in formal analogy to a hazard function or force of mortality in survival analysis  provides comparisons of functional trajectories based on their exceedance behavior. By integrating these concepts with Fr\'echet regression, the proposed methodology facilitates the analysis of covariate effects on exceedance patterns.

Promising future research directions include an extension to multivariate (vector) functional data, which could illuminate interactions among interdependent exceedance events. Another avenue lies in extending the approach to functional data with a multivariate time domain. A related topic is to  adapt the methodology for spatial and spatiotemporal analysis, where exceedances of random fields are of interest.  Another future extension of interest is to incorporate uncertainty quantification.

\small 
\bibliographystyle{abbrvnat}
\bibliography{citation}

\end{document}